\documentclass[onecollarge]{svjour2}       

\smartqed  

\usepackage{mathptmx}      

\usepackage{amsmath,amssymb}
\usepackage{bbm}

\newcommand\N{{\mathbbm{N}}}
\newcommand\e{{\mathrm{e}}}

\journalname{}

\begin{document}

\title{Recursion relations for the partition function of the two-dimensional Ising model}


\author{Michael Kastner       
}


\institute{Michael Kastner \at Physikalisches Institut, Universit\"at Bayreuth, 95440 Bayreuth, Germany;\\
              \email{Michael.Kastner@uni-bayreuth.de}           
}

\date{Received: date / Accepted: date}

\maketitle

\begin{abstract}
The partition function of the two-dimensional Ising model on a square lattice with nearest-neighbour interactions and periodic boundary conditions is investigated. Kaufman [Phys.\ Rev.\ {\bfseries 76}, 1232--1243 (1949)] gave a solution for this function consisting of four summands. The summands are rewritten as functions of a low-temperature expansion variable, resulting in polynomials with integer coefficients. Considering these polynomials for system sizes $2^m\times 2^n$ ($m,n\in\N$), a variety of recursion relations in $m,n$ are found. The recursions reveal a rich structure of the partition function and can be employed to render the computer algebra calculation of the microcanonical partition function more efficient.
\end{abstract}

\section{Introduction}
\label{sec:intro}
Every scientific community has its favourite model system, and the role which is played in genetics by {\em drosophila melanogaster}\/ can be attributed to the Ising model in the statistical physics of phase transitions. It is arguably the best studied model in the field, and advances in the understanding of the Ising model have significantly advantaged progress in the study of phase transitions in general.

Proposed by W.~Lenz in 1920, the one-dimensional version of this lattice model with nearest-neighbour interactions was solved by his student E.~Ising \cite{Ising:25} in 1925. Ising showed the absence of a phase transition in the one-dimensional case, and this result, enforced by Ising's claim that likewise no phase transition should take place in higher dimensions, discouraged further studies for a while. It was Peierls' proof \cite{Peierls:36,Griffiths:64} of the existence of a phase transition in the two-dimensional Ising model on a square lattice which rekindled the interest, culminating in Onsager's exact solution \cite{Onsager:44} for the free energy density, published in 1944 (see \cite{Niss:04} for an account of the early history of the Ising model). 

The above cited articles all focus on properties of the Ising model in the thermodynamic limit, in which the number of lattice sites goes to infinity. In 1949, B.~Kaufman published an extended version of Onsager's solution, reporting an exact expression also for the canonical partition function of the two-dimensional Ising model on a finite square lattice with periodic boundary conditions \cite{Kaufman:49}. Almost 50 years later, P.~D.~Beale \cite{Beale:96} employed a computer algebra program to calculate from Kaufman's result, for modest lattice sizes of up to $32\times32$ sites, the exact microcanonical partition function for the energy, a quantity particularly useful when testing the performance of computer simulation algorithms. The microcanonical partition function, taking on values from the non-negative integers $\N_0$, is given by the number of configurations compatible with a certain value of the energy. From Beale's algorithm, these integer numbers are computed from sums and products of {\em irrational} numbers. This does not appear to be the straightforward way to calculate an integer number, and it was this observation which made me start investigating the structure of the finite system partition function of the two-dimensional Ising model.

When working on this problem, I came across beautiful numbers and rich structures, mainly in the form of recursion relations between polynomials with integer coefficients. Although the solution obtained is not complete (in a sense which will become clear later), I would like to share at least the problem, its beauty, and a partial solution with the scientific community. Apart from the esthetic appeal, the study can have the practical application of improving the performance of Beale's algorithm.

\section{The two-dimensional Ising model}
\label{sec:ising}
Consider a finite two-dimensional square lattice
\begin{equation}
{\mathcal L}_{M,N}=\left\{1,\dotsc,M\right\}\times\left\{1,\dotsc,N\right\},
\end{equation}
consisting of $MN$ lattice sites. Periodic boundary conditions are assumed, and to each lattice site $i\in{\mathcal L}_{M,N}$ a classical spin variable $\sigma_i\in\left\{-1,+1\right\}$ is associated. Then the two-dimensional Ising model (without external magnetic field) is defined by the Hamiltonian function
\begin{equation}\label{eq:hamiltonian}
{\mathcal H}(\sigma)=-J\sum_{\langle i,j\rangle}\sigma_i \sigma_j,
\end{equation}
mapping each configuration $\sigma=\left\{\sigma_i\right\}_{i\in{\mathcal L}_{M,N}}$ from the configuration space
\begin{equation}
\Gamma_{MN}=\left\{-1,+1\right\}^{\times MN}
\end{equation}
onto $J$ times the even integer numbers
\begin{equation}\label{eq:spectrum}
\left\{0,2,4\dotsc,\pm2MN-6,\pm2MN-4,\pm2MN \right\}.
\end{equation}
The positive constant $J$ determines the coupling strength and the angle brackets $\left\langle\cdot,\cdot\right\rangle$ in \eqref{eq:hamiltonian} denote a summation over all pairs of nearest neighbours on the lattice ${\mathcal L}_{M,N}$ (each pair counted once).

\section{Kaufman's expression for the canonical partition function}
\label{sec:partition}
The canonical partition function is given by
\begin{equation}
{\mathcal Z}_{M,N}^{\phantom{{(i)}}}(K)=\sum_{\sigma\in\Gamma_{MN}}\e^{-\beta{\mathcal H}(\sigma)},
\end{equation}
where $K=2J\beta$ and $\beta=\frac{1}{k_B T}$ is the inverse of temperature $T$ times Boltzmann's constant $k_B$. The $K$-dependence of ${\mathcal Z}_{M,N}^{\phantom{{(i)}}}$ will be suppressed in the following to lighten the notation. Kaufman \cite{Kaufman:49} showed that the canonical partition function of the two-dimensional Ising model can be written  in the form
\begin{equation}\label{eq:Z_Kaufman}
2\, {\mathcal Z}_{M,N}^{\phantom{{(i)}}}={\mathcal Z}_{M,N}^{(1)}+{\mathcal Z}_{M,N}^{(2)}+{\mathcal Z}_{M,N}^{(3)}+{\mathcal Z}_{M,N}^{(4)},
\end{equation}
where
\begin{subequations}
\begin{align}
{\mathcal Z}_{M,N}^{(1)}& =\left(2\sinh K\right)^{MN/2}\,\prod_{k=0}^{N-1} 2\cosh\left(\frac{M}{2}\gamma_{2k+1}\right),\\
{\mathcal Z}_{M,N}^{(2)}& =\left(2\sinh K\right)^{MN/2}\,\prod_{k=0}^{N-1} 2\sinh\left(\frac{M}{2}\gamma_{2k+1}\right),\\
{\mathcal Z}_{M,N}^{(3)}& =\left(2\sinh K\right)^{MN/2}\,\prod_{k=0}^{N-1} 2\cosh\left(\frac{M}{2}\gamma_{2k}\right),\\
{\mathcal Z}_{M,N}^{(4)}& =\left(2\sinh K\right)^{MN/2}\,\prod_{k=0}^{N-1} 2\sinh\left(\frac{M}{2}\gamma_{2k}\right).
\end{align}
\end{subequations}
For $0<k<2N$ the quantity $\gamma_k$ is defined as the positive root of
\begin{subequations}\label{eq:gamma}
\begin{equation}\label{eq:gamma_k}
\cosh\left(\gamma_k\right) = \frac{\cosh^2 K}{\sinh K} - \cos\left(\frac{\pi k}{N}\right),
\end{equation}
whereas for $k=0$ we have
\begin{equation}
\e^{\gamma_0}=\e^K \tanh\left(K/2\right).
\end{equation}
\end{subequations}
The invariance of ${\mathcal Z}_{M,N}^{\phantom{{(i)}}}$ upon interchange of the parameters $M$ and $N$, although evident from the definition of the Ising model, is not quite obvious in the above expressions.

As a consequence of the equidistant energy values \eqref{eq:spectrum} the canonical partition function can be expressed as a low-temperature series of the form
\begin{equation}\label{eq:Z_lt}
{\mathcal Z}_{M,N}^{\phantom{{(i)}}}=x^{-MN}\sum_{k=0}^{MN} g_k x^{2k},
\end{equation}
where 
\begin{equation}
x=\e^{-K},
\end{equation} 
is the low-temperature expansion variable. This is exactly what is done by the computer algebra program at the core of Beale's work \cite{Beale:96}: For given $M$ and $N$, the solution specified by equations \eqref{eq:Z_Kaufman}\nobreakdash--\eqref{eq:gamma} is cast into the form \eqref{eq:Z_lt}. From this result the microcanonical partition function $g_k$, specifying the number of configurations with energy $-2J\left(MN-2k\right)$, can be read off. The $g_k$ are non-negative integers, but their computation from equations \eqref{eq:Z_Kaufman}\nobreakdash--\eqref{eq:gamma} involves sums and products of irrational numbers, stemming essentially from the cosine in \eqref{eq:gamma_k}.

\section{Low-temperature expansions of the ${\mathcal Z}_{M,N}^{(i)}$}
\label{sec:expansion}
Since low-temperature expansions show the appealing property of having integer expansion coefficients, we continue working in this representation. One can similarly express the four summands ${\mathcal Z}_{M,N}^{(i)}$ ($i=1,2,3,4$) in \eqref{eq:Z_Kaufman} in the form of a low-temperature expansion,
\begin{equation}\label{eq:Zi_lt}
{\mathcal Z}_{M,N}^{(i)}=x^{-MN}\sum_{k=0}^{MN} g_k^{(i)} x^{2k},
\end{equation}
with expansion coefficients $g_k^{(i)}$. From Kaufman's solution \eqref{eq:Z_Kaufman}\nobreakdash--\eqref{eq:gamma} or from Beale's computer algebra program in \cite{Beale:96} one obtains for example
\begin{subequations}
\begin{align}
x^{4}{\mathcal Z}_{2,2}^{(1)} =& \big(1+x^2\big)^4,\\
x^{16}{\mathcal Z}_{4,4}^{(1)} =& \big(1+x^2\big)^8\big(1-4x^2+22x^4-4x^6+x^8\big)^2,\\
x^{36}{\mathcal Z}_{6,6}^{(1)} =& \big(1+x^2\big)^{12}\big(1-8x^2+30x^4-8x^6+x^8\big)^2\big(1+x^2+12x^4+x^6+x^8\big)^4,\\
x^{64}{\mathcal Z}_{8,8}^{(1)} =& \big(1+x^2\big)^{16}\big(1+20x^4+32x^6+150x^8+32x^{10}+20x^{12}+x^{16}\big)^4\notag\\ & \times\big(1-8x^2+28x^4-56x^6+326x^8-56x^{10}+28x^{12}-8x^{14}+x^{16}\big)^2.
\end{align}
\end{subequations}
Remarkably, not only the $g_k$ from the series \eqref{eq:Z_lt}, but already the $g_k^{(i)}$ in \eqref{eq:Zi_lt} are integer numbers (not necessarily non-negative, however). So the ${\mathcal Z}_{M,N}^{(i)}$ are essentially polynomials in $x$ with integer coefficients. The complexity of these polynomials can be reduced significantly by the transformations of variables
\begin{equation}\label{eq:trafo}
y=\left(\frac{1-x^2}{2x}\right)^2 \qquad\text{and}\qquad z=\frac{4(1+y)^2}{y},
\end{equation}
yielding for example
\begin{subequations}\label{eq:Zcalz}
\begin{align}
{\mathcal Z}_{2,2}^{(1)} =& \left(4y\right)^{1} z,\label{eq:Z1z22}\\
{\mathcal Z}_{4,4}^{(1)} =& \left(4y\right)^{4} z^2\left(8-z\right)^2,\label{eq:Z1z44}\\
{\mathcal Z}_{6,6}^{(1)} =& \left(4y\right)^{9} z^3\left(12-z\right)^2\left(3-z\right)^4,\\
{\mathcal Z}_{8,8}^{(1)} =& \left(4y\right)^{16} z^4\left(32-16z+z^2\right)^2 \left(8-8z+z^2\right)^4\label{eq:Z1z88}.
\end{align}
\end{subequations}
In the following we will establish recursion relations between the canonical partition functions ${\mathcal Z}_{M,N}^{\phantom{{(i)}}}$ and ${\mathcal Z}_{2M,N}^{\phantom{{(i)}}}$ or ${\mathcal Z}_{M,2N}^{\phantom{{(i)}}}$, respectively. To this purpose, and for further simplification, we restrict ourselves to the Ising model on lattices ${\mathcal L}_{2^m,2^n}$ where $m,n\in\N$. Then it is convenient to introduce the notation
\begin{equation}
Z_{m,n}^{(i)}=\left(4y\right)^{-2^{m+n-2}}{\mathcal Z}_{2^m,2^n}^{(i)},
\end{equation}
yielding for example
\begin{subequations}\label{eq:Zz}
\begin{align}
Z_{1,1}^{(1)} =& z,\\
Z_{2,2}^{(1)} =& z^2\left(8-z\right)^2,\\
Z_{3,3}^{(1)} =& z^4\left(32-16z+z^2\right)^2 \left(8-8z+z^2\right)^4,\\
Z_{4,4}^{(1)} =& z^8\left(512-1024z+320z^2-32z^3+z^4\right)^2 \left(8-32z+40z^2-16z^3+z^4\right)^4\notag\\ &\times\left(8-96z+72z^2-16z^3+z^4\right)^4 \left(32-128z+80z^2-16z^3+z^4\right)^4,
\end{align}
\end{subequations}
(compare with equations \eqref{eq:Z1z22}, \eqref{eq:Z1z44}, and \eqref{eq:Z1z88}). The rest of this article is devoted to the study of relations among the $Z_{m,n}^{(i)}$. Instead of having tried to obtain these relations from the general equations \eqref{eq:Z_Kaufman}\nobreakdash--\eqref{eq:gamma}, the $Z_{m,n}^{(i)}$ were calculated by means of a computer algebra program for several values of $m$ and $n$ from which the relations were ``read off''. It is for this reason that no derivations are given in sections \ref{sec:symmetry}\nobreakdash--\ref{sec:relationsZmn1}. The validity of the relations has been verified for values of $m,n$ up to $7$, corresponding to linear lattice sizes up to $128$ sites. The present article is supplemented by a {\sc Mathematica} notebook available at http://www.phy.uni-\linebreak bayreuth.de/~btp308/Zmn.tar.gz containing a data set and some routines which allow to construct and handle the $Z_{m,n}^{(i)}$, given as polynomials in $z$.

\section{Symmetry relations for the $Z_{m,n}^{(i)}$}
\label{sec:symmetry}
For any given $m,n\in\N$, the symmetry relations
\begin{equation}
Z^{(1)}_{m,n}=Z^{(1)}_{n,m},\qquad Z^{(4)}_{m,n}=Z^{(4)}_{n,m},\qquad\text{and}\qquad Z^{(2)}_{m,n}=Z^{(3)}_{n,m}
\end{equation}
hold. In combination, these relations imply the invariance
\begin{equation}
{\mathcal Z}_{2^m,2^n}={\mathcal Z}_{2^n,2^m}
\end{equation}
of the canonical partition function.

\section{Relations between the $Z_{m,n}^{(i)}$}
\label{sec:relationsZmni}
The $Z_{m,n}^{(i)}$ are polynomials in $z$ of degree $2^{m+n-2}$. As may already be suggested by \eqref{eq:Zz}, these polynomials typically factorize into several polynomials of lower degree. Many of these factors appear repeatedly in the various $Z_{m,n}^{(i)}$, and, for all $m,n\geqslant2$, the relations
\begin{subequations}\label{eq:Zrecursions}
\begin{align}
Z^{(2)}_{m,n} & = Z^{(1)}_{m-1,n}Z^{(2)}_{m-1,n},\\
Z^{(3)}_{m,n} & = Z^{(1)}_{m,n-1}Z^{(3)}_{m,n-1} = Z^{(2)}_{1,m}\prod_{k=1}^{n-1} Z^{(1)}_{m,k},\\
Z^{(4)}_{m,n} & = Z^{(2)}_{m,n-1}Z^{(4)}_{m,n-1} = Z^{(4)}_{1,m}\prod_{k=1}^{n-1} Z^{(2)}_{m,k},
\end{align}
\end{subequations}
can be read off. To compute the various $Z_{m,n}^{(i)}$ ($i=2,3,4$) from these relations, the $Z_{1,n}^{(i)}$ ($i=2,3,4$) are needed as starting values, for which we find
\begin{subequations}
\begin{align}
Z^{(2)}_{1,n}&=b_n,\\
Z^{(3)}_{1,n}&=b_1\prod_{k=1}^{n-1}a_k^2,\\
Z^{(4)}_{1,n}&=\frac{4(y^2-1)}{y}\prod_{k=1}^{n-1}b_k.
\end{align}
\end{subequations}
The $a_n$ are given by the recursion relation
\begin{subequations}\label{eq:an}
\begin{align}
a_1&=\sqrt{z},\\
a_n&=a_{n-1}^2-2 \qquad\text{for $n\geqslant2$},
\end{align}
\end{subequations}
yielding a sequence whose first few terms read
\begin{subequations}\label{eq:an_sequence}
\begin{align}
a_1 &= \sqrt{z},\\
a_2 &= -2+z,\\
a_3 &= 2-4z+z^2,\\
a_4 &= 2-16z+20z^2-8z^3+z^4,\\
a_5 &= 2-64z+336z^2-672z^3+660z^4-352z^5+104z^6-16z^7+z^8.
\end{align}
\end{subequations}
The $b_n$ are determined by
\begin{subequations}\label{eq:bn}
\begin{align}
b_1&=-4+z,\\
b_n&=(b_{n-1}+2)^2-8\beta_{n-1} \qquad\text{for $n\geqslant2$},
\end{align}
\end{subequations}
where
\begin{subequations}
\begin{align}
\beta_1 &=z,\\
\beta_n &=\left(\beta_{n-1}-2+4^{n-1} z\prod_{k=2}^{n-2}\beta_k\right)^2 \qquad\text{for $n\geqslant2$},
\end{align}
\end{subequations}
and the resulting sequence starts with
\begin{subequations}\label{eq:bn_sequence}
\begin{align}
b_1 &= -4+z,\\
b_2 &= 4-12z+z^2,\\
b_3 &= 4-176z+148z^2-24z^3+z^4,\\
b_4 &= 4-2752z+29520z^2-52704z^3+30356z^4-7456z^5+872z^6-48z^7+z^8.
\end{align}
\end{subequations}
The results of this section effectively reduce the calculation of the $Z_{m,n}^{(i)}$ for {\em any}\/ $i,m,n$ to the computation of the $Z_{m,n}^{(1)}$.

\section{Recursion relations for the $Z_{m,n}^{(1)}$}
\label{sec:relationsZmn1}
For an efficient computation it would likewise be desirable to find recursion relations for the $Z_{m,n}^{(1)}$ for different $m,n$. For $m=1,2$ (or, equivalently, $n=1,2$), $Z_{m,n}^{(1)}$ can be written as
\begin{equation}
Z^{(1)}_{1,n}=a_n^2\qquad\text{and}\qquad Z^{(1)}_{2,n}=c_n^2
\end{equation}
with the $a_n$ as given in \eqref{eq:an}. The $c_n$ are, again, given by recursion relations,
\begin{subequations}
\begin{align}
c_1&=-2+z
,\\
c_2&=
z(z-8),\\
c_n&=(c_{n-1}+2)^2-\frac{1}{2}(c_{n-1}-c_{n-2}^2+4)^2\qquad\text{for $n\geqslant3$},
\end{align}
\end{subequations}
yielding a sequence that starts with
\begin{subequations}\label{eq:cn_sequence}
\begin{align}
c_1 &= -2+z,\\
c_2 &= z \left(-8+z\right),\\
c_3 &= 4-32z+60z^2-16z^3+z^4,\\
c_4 &= 4-128z+1264z^2-4160z^3+4628z^4-1984z^5+376z^6-32z^7+z^8.
\end{align}
\end{subequations}
The next sequence of interest would be $\big\{ Z_{3,n}^{(1)}\big\}_{n\in\N}$, starting as
\begin{subequations}\label{eq:Z1_3n}
\begin{align}
Z_{3,1}^{(1)} =& \left(2-4z+z^2\right)^2,\\
Z_{3,2}^{(1)} =& \left(4-32z+60z^2-16z^3+z^4\right)^2,\\
Z_{3,3}^{(1)} =& z^4\left(32-16z+z^2\right)^2 \left(8-8z+z^2\right)^4,\\
Z_{3,4}^{(1)} =& \big(4-352z+3424z^2-6976z^3+5436z^4-2000z^5+368z^6-32z^7+z^8\big)^2\notag\\
	      \times& \big(4-160z+1376z^2-3648z^3+3900z^4-1840z^5+368z^6-32z^7+z^8\big)^2,\\
Z_{3,5}^{(1)} =& \big(4-512z+23936z^2-504128z^3+4799728z^4-20395520z^5\notag\\
	       &+44112832z^6-55088640z^7+42942996z^8-21874432z^9+7464768z^{10}\notag\\
	       &-1722784z^{11}+267640z^{12}-27392z^{13}+1760z^{14}-64z^{15}+z^{16}\big)^2\notag\\
	      \times& \big(4-1536z+149888z^2-1832128z^3+9714928z^4-28345856z^5\notag\\
	       &+50377664z^6-57468672z^7+43287060z^8-21857536z^9+7456064z^{10}\notag\\
	       &-1722208z^{11}+267640z^{12}-27392z^{13}+1760z^{14}-64z^{15}+z^{16}\big)^2,
\end{align}
\end{subequations}
(for more terms of this sequence see the accompanying {\sc Mathematica} notebook at http://www.phy.uni-\linebreak bayreuth.de/~btp308/Zmn.tar.gz). I was not able to obtain recursion relations (in $n$) for this sequence, neither for the other $Z_{m,n}^{(1)}$ with $m\geqslant4$. Similarly, recursions for $Z_{n,n}^{(1)}$, i.\,e., for the Ising model on square lattices of linear size $2^n$, would be desirable.

\section{Summary}
\label{sec:summary}
We investigated Kaufman's solution \cite{Kaufman:49} of the partition function of the two-di\-men\-sion\-al Ising model on an $M\times N$ square lattice with nearest-neighbour interactions and periodic boundary conditions. The four summands ${\mathcal Z}_{M,N}^{(i)}$ ($i=1,2,3,4$) contributing to this solution were considered separately. Rewriting these summands as functions of the low-temperature expansion variable $x=\exp\left(-2J\beta\right)$, polynomials with integer coefficients were obtained. The polynomials can be simplified significantly by the change of variables defined in \eqref{eq:trafo}. Restricting the analysis to lattices of sizes $2^m\times 2^n$ ($m,n\in\N$), a variety of recursion relations in $m,n$ were found for the resulting simplified polynomials, and hence for the ${\mathcal Z}_{M,N}^{(i)}$. These recursion relations reduce the calculation of the $Z_{m,n}^{(i)}$ for any $i,m,n$ to the computation of $Z_{m,n}^{(1)}$. For $m=1,2$, the $Z_{m,n}^{(1)}$ can again be computed efficiently from recursion relations for arbitrary $n$. The search for similar recursions for $Z_{m,n}^{(1)}$ when $m\geqslant3$ was unsuccessful and remains an open problem. 

The results obtained are of interest for different reasons. First, the recursion relations can be employed to render more efficient the (computer algebra) calculation of the microcanonical partition function in the spirit of \cite{Beale:96}. With the results obtained so far, the gain in efficiency is still modest; if, however, recursion relations like the ones indicated as open problems in section \ref{sec:relationsZmn1} could be found, the efficiency of the computation would increase drastically.

Second, the results reported in this article reveal a fascinatingly rich structure of the partition function. In my opinion it is in particular the reformulation of the ${\mathcal Z}_{M,N}^{(i)}$ as polynomials in $z$ with integer coefficients which contributes significantly to the esthetic appeal of the problem discussed in this article. My hope would be that the results presented attract the interest of someone capable of making further progress in the direction pointed out.

\bibliographystyle{spmpsci}      
\bibliography{2dIsing}   

\end{document}